\documentclass[twocolumn,secnumarabic,amssymb, nobibnotes, aps, prd]{revtex4}
\usepackage{graphicx}

\begin{document}
\title{The quantum challenge on mesoscopic level}
\author{A.V. Nikulov}
%\email[]{nikulov@ipmt-hpm.ac.ru}
\affiliation{Institute of Microelectronics Technology and High Purity Materials, Russian Academy of Sciences, 142432 Chernogolovka, Moscow District, RUSSIA.} %nikulov@ipmt-hpm.ac.ru
%\date{}
\begin{abstract} Most experts reject the quantum potential introduced by David Bohm in 1952. But it is impossible to describe some quantum mesoscopic phenomena observed in nanostructures without a quantum force.
 \end{abstract}

\maketitle

\narrowtext

\section*{Introduction}

The Lecture by one of the Nobel Prize winners presented at the Eighth International Symposium "Frontiers of Fundamental Physics"  [1] called "What is Quantum Mechanics?" This question  may seem strange for many scientists who studied and used formalism of quantum mechanics. We can describe numerous phenomena of atomic physics, solid state physics, mesoscopic physics and many others using the so-called Copenhagen formalism. But experts understand that quantum mechanics {\it not yet based on a generally accepted conceptual foundation} [2]. Albert Einstein has always regarded the Copenhagen interpretation of the quantum theory as incomplete and has always believed  that, even at the quantum level, there must exist precisely definable variables determining the actual behavior of each individual system. This point of view known as "realism" is the basis of hidden-variable theories [3]. Bell's theorem [4], and the experiments [5] based on it, have provided convincing evidence that a local hidden-variable theory will never be able to account for the full rang of quantum phenomena. But the very first hidden-variable theory proposed by David Bohm in 1952 [6] was radically nonlocal. In order to save the principle of realism he must conclude that nonlocality can be real. The assumption on a real nonlocality seems so strange that most experts prefer to renounce realism [3]. Nevertheless one should not reject without hesitation the nonlocal quantum potential introduced by Bohm [6]. The well known Aharonov-Bohm effect [7] can bear a relation to this problem [8]. The nonlocal momentum transfer [9] observed in the Aharonov-Bohm double-slit experiment can be interpreted in terms of complementarity [3]. But such interpretation can be not possible in some cases of the Aharonov-Bohm effect observed in nanostructures. This problem is urgent because of numerous Aharonov-Bohm phenomena observed in semiconductor [10], normal metal [11] and superconductor [12] mesoscopic systems. 
\begin{figure}[b]
\includegraphics{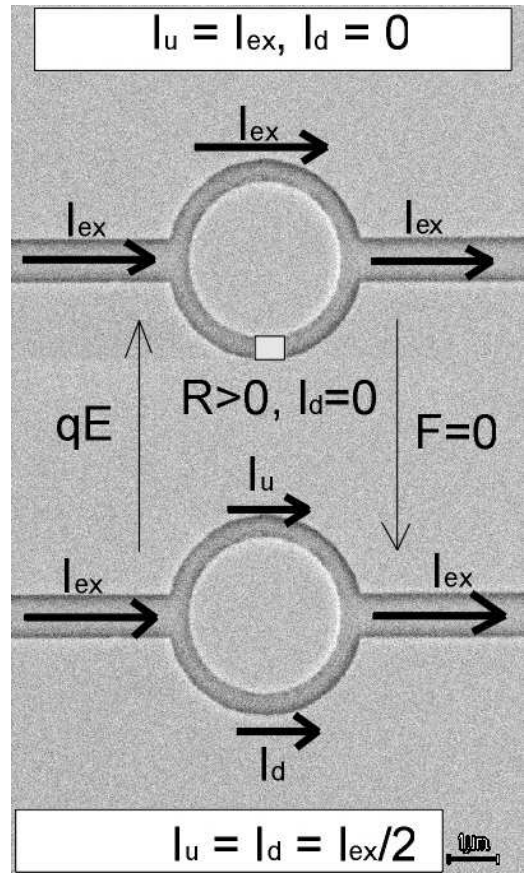}
\caption{\label{fig:epsart} Example of momentum change induced by switching between states with different connectivity of the wave function.}
\end{figure}

\section {Quantum force}
The Bohm's theory is based on the Schrodinger equation. Bohm has divided it into two, one for the amplitude $|\Psi |$ of wave function $\Psi = |\Psi | \exp i\varphi $ and the other for the phase $\varphi $ [6]. When the wave function describes a probability, as for example in the double-slit experiment, it is enough easy to reject the Bohm's quantum potential, inadmissible for most experts, and to repudiate realism. But this renunciation of realism can be impossible when the wave function describing a density, for example superconducting pairs. There is an experimental challenge to the Copenhagen interpretation connected with observations of the persistent current $I_{p} \neq 0$ in semiconductor [13], normal metal [11] and superconductor [14] rings with non-zero resistance $R_{l} > 0$. According to classical physics a circular direct current $I$ is possible at $R_{l} > 0$ only at a non-zero Faraday's voltage $d\Phi/dt \neq 0$ when the force balance $IR_{l} = -d\Phi/dt $ takes place. The observations $I_{p} \neq 0$ at $R_{l} > 0$ and $d\Phi/dt = 0$ violate this balance. In order to restore it a quantum force was introduced in [15]. 

\begin{figure}
\includegraphics{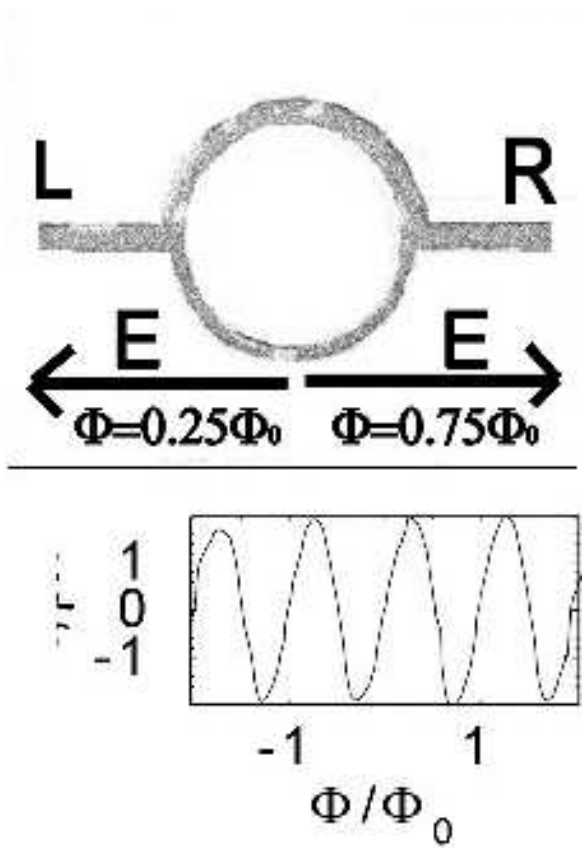}
\caption{\label{fig:epsart} Aluminum asymmetric loop with radius $2 \ \mu m$ and the quantum oscillations of the dc potential difference $V(\Phi/\Phi_{0})$ ($\mu V$) observed on such loop. }
\end{figure}

The quantum force introduced in [15] describes a change of momentum $p$ and velocity $v_{s}$ of superconducting pairs because of the Bohr's quantization $\oint_{l}dl p = \oint_{l}dl \hbar \bigtriangledown \varphi  =  \oint_{l}dl (mv_{s} + qA) = m\oint_{l}dlv + q\Phi =  n2\pi \hbar $ at closing of wave function: $ v_{s} = 0$ and $\oint_{l}dl p =  q\Phi$ at unclosed wave function whereas $\oint_{l}dl p = n2\pi \hbar $ and $\oint_{l}dlv = (1/m)(n2\pi \hbar - q\Phi) = (2\pi \hbar /m)(n - \Phi/\Phi_{0})$ at closed wave function. $\Phi_{0} = 2\pi \hbar /q = \pi \hbar /e$ is the flux quantum; $q =2e$ is the charge of superconducting pair. The necessity of the quantum force is obvious from the example shown on Fig.1. The external current $I_{ex}$ flows on the top semi-ring of the superconducting loop, $I_{u} = I_{ex}$, $I_{d} = 0$, when a segment of the lower one in the normal state with a resistance $R > 0$. But the currents should be equal in the both semi-rings, $I_{u} = I_{ex}/2$, $I_{d} = I_{ex}/2$, because of the Bohr's quantization $\oint_{l}dlv = l_{u}v_{u} - l_{d}v_{d} \propto  (n - \Phi/\Phi_{0}) = 0$ after switching of this segment into superconducting state. This change of the current values occurs without any electric field and it could not occur in an ideal conductor at switching from $R > 0$ to $R = 0$.  

\section{Intrinsic breach of symmetry}
There is an important difference between atomic and mesoscopic levels [16]: it is impossible to realize the switching between states with different connectivity of the wave function on the atomic level whereas it is possible in real mesoscopic structures. A potential difference with a dc component $V_{dc} = L \omega I_{ex}/2$ should be observed both on top and lower semi-rings when a segment only in the lower semi-ring, Fig.1, is switched with a frequency $\omega < R/L$ between superconducting and normal state with $R > 0$. $L$ is the semi-ring inductance. The dc potential difference $V_{dc} = L \omega I_{p}(\Phi/\Phi_{0})$ can be observed [17] also without any dc external current because of the persistent current $ I_{p}$, the value and sign of which is periodical function of magnetic flux $\Phi$ inside the ring with period equal the flux quantum $\Phi_{0}$. Such quantum oscillations of the dc voltage $V_{dc}(\Phi/\Phi_{0}) \propto I_{p}(\Phi/\Phi_{0})$ were observed on segments of asymmetric aluminum ring [12]. The periodical change of the dc electric field $E(\Phi/\Phi_{0}) = -\bigtriangledown V(\Phi/\Phi_{0})$ direction with the scalar value $\Phi$ observed in this experiment gives experimental evidence of intrinsic breach of right-left symmetry [18].  

This result has fundamental importance and should be connected with the necessity of the quantum force. There was a logical difficulty in the Bohr's model until electron considered as a particle having a velocity since it was impossible to answer on the question: "What direction has this velocity?" The uncertainty relation and the wave quantum mechanics have overcame this difficulty. Albert Einstein considered it as weakness: {\it The weakness of the theory lies … in the fact, that it leaves time and direction of the elementary process to "chance"} (the citation from  [2]). Thanks to this "weakness" the Bohr's quantization does not violate symmetry between opposite directions on the atomic level. But we see experimental evidence of the intrinsic breach of symmetry because of the Bohr's quantization in nanostructures. 

\section*{Acknowledgement}
This work has been supported by a grant 04-02-17068 of the  Russian Foundation of Basic Research.

\end{document}